\documentclass[author]{nrc1}
\usepackage{amsmath}
\usepackage{amsfonts}
\usepackage{amssymb}
\usepackage[ansinew]{inputenc} 
\usepackage{graphicx,color}
\usepackage{subfigure} 

\usepackage{lineno}
\usepackage{booktabs}
\usepackage{algorithm2e}

\begin{document}
\title{Impact of Global Data Assimilation System  atmospheric models on astroparticle showers}

\author[Jennifer Grisales-Casadiegos]{J. Grisales-Casadiegos}
\address[1]{Escuela de F\'isica, Facultad de Ciencias, Universidad Industrial de Santander, Bucaramanga, Bucaramanga, 680002, Colombia. jennifer.grisales@saber.uis.edu.co}
\author[Christian Sarmiento-Cano]{C. Sarmiento-Cano}
\address[2]{Instituto de Tecnolog\'ia Sabato, Universidad Nacional de San Mart\'in,  Centro At\'omico Constituyentes, and Instituto en Tecnolog\'ias de Detecci\'on y Astropart\'iculas, Buenos Aires, Argentina}
\author[Luis A. N\'u\~nez]{L.A. N\'u\~nez and the LAGO Collaboration}
\address[4]{Escuela de F\'isica, Facultad de Ciencias, Universidad Industrial de Santander, Bucaramanga, 680002, Colombia and Departamento de F\'{\i}sica, Facultad de Ciencias, Universidad de Los Andes, M\'{e}rida 5101, Venezuela}
\shortauthor{Grisales-Casadiegos, Sarmiento-Cano,  N\'u\~nez and the LAGO Collaboration}
\maketitle

\begin{abstract}
We present a methodology to simulate the impact of the atmospheric models in the background particle flux on ground detectors using the Global Data Assimilation System. The methodology was within the ARTI simulation framework developed by the Latin American Giant Observatory Collaboration. The ground level secondary flux simulations were performed with a tropical climate at the city of Bucaramanga, Colombia. To validate our methodology, we built monthly profiles over Malarg\"ue between 2006 and 2011, comparing the maximum atmospheric depth, X$_\mathrm{max}$, with those calculated with the Auger atmospheric option in CORSIKA. The results show significant differences between the predefined CORSIKA atmospheres and their corresponding Global Data Assimilation System atmospheric profiles.
\end{abstract}

\section{Introduction}
The interaction of cosmic rays with the nuclei of atomic elements in the atmosphere produces a cascade of particles: the extensive air shower (EAS). These cascades, measured at the Earth's surface, result from an intricate convolution of physical phenomena: including dispersion, decay and absorption of high energy particles crossing the atmosphere. For energies below $10^{12}$~eV, the primaries are modulated by the solar wind. Thus they interact with the interplanetary and geomagnetic fields, resulting in what is known as Space Weather Physics. An excellent overview of this exciting field can be found in ~\cite{UsoskinEtal2002, BothmerDaglis2007, BarnardEtal2011, Potgieter2013} and references therein. 

The Latin American Giant Observatory (LAGO) is an astroparticle observatory with water Cherenkov detectors (WCD) distributed from Mexico to Antarctica, covering a broad range of geomagnetic rigidity cutoffs and atmospheric depths (from sea level up to more than 5000~m.a.s.l.~\cite{SidelnikAsorey2017}). LAGO has developed a Space Weather program~\cite{AsoreyEtal2015B} to explore the influence of space climate on the cosmic ray flux at ground level. This research program has developed a precise simulation scheme, taking into account the geomagnetic corrections, providing an estimation of the response of a Water Cherenkov Detector (WCD) to the impinging particle flux at any geographic place on Earth's surface~\cite{AsoreyNunezSuarez2018}.  The simulation pipeline of computing algorithms considers three factors with different spatial and time scales: the geomagnetic effects, the development of the extensive air showers in the atmosphere, and the detector response at ground level.

In this work we use GDASTOOL --a python routine based on the Global Data Assimilation System (GDAS)~\cite{Will2014}-- as part of a methodology use in creating monthly atmospheric profiles for any location. This routine is available in the latest versions of CORSIKA (Cosmic Ray Simulations for KASKADE~\footnote{KASKADE: KArlsruhe Shower Core and Array DEtector}) and allows us to obtain a specific atmosphere model for a particular day, time and geographic location. GDASTOOL helps us visualize the impact that detailed atmospheric and climate models have on the cosmic ray flux at the Earth's surface. 

We show the usefulness these profiles have in reproducing the atmospheric conditions in which an EAS develops. Also, by applying the described methodology to the city of Bucaramanga-Colombia, we determined the influence of the monthly atmospheric models on the flux of secondary particles arriving at the LAGO WCD detectors. With this insight, we discuss the relevance of implementing GDAS at any geographical location.

This work is organized as follows: In the next section, we briefly describe the Latin American Giant Observatory and present the structure of the ARTI computational framework. In section~\ref{sec:GDAS}, we discuss the methodology used to build the atmospheric models and the GDAS implementation at the Pierre Auger Observatory. Then, in section~\ref{sec:FluxDependence} we present the results obtained by comparing the background flux reaching the surface using the standard tropical atmosphere versus the GDAS atmospheres. Finally we will end with some remarks displayed in section~\ref{sec:FinalRemarks}. 

\section{LAGO and the ARTI simulation framework}
\label{LAGOARTI}
The Latin American Giant Observatory is an extended observatory on a continental scale,  promoting training and research in astroparticle physics in Latin America. Its three main research areas are space weather phenomena, search for gamma rays bursts at high altitude sites and background radiation at ground level.

The LAGO WCD detection network extends over several sites at different latitudes and altitudes, covering various geomagnetic rigidity and atmospheric depths. The low-cost LAGO WCDs have proven detection reliability and efficiency for both electromagnetic and muon EAS components~\cite{SidelnikAsorey2017}.

The LAGO Space Weather Program studies the relation between the variation in the flux of secondary particles at ground level with the heliospheric modulation of  Galactic Cosmic Rays (GCR). Through the multi-spectral analysis technique~\cite{AsoreyEtal2015B}, LAGO provides detailed information of the temporal evolution of the secondary flux at different geomagnetic locations, contributing to a better understanding better in the disturbances produced by different space weather phenomena~\cite{Suarezduran2019en}.

The estimation of the expected flux of secondaries at any detector of the LAGO network is based on a detailed and realistic simulation considering all possible sources of flux variation. This complex approach involving processes occurring at different spatial and time scales, can be depicted as:

\begin{center}
\begin{table}[htb!]
\begin{tabular}{llllll}
GCR Flux & $\xrightarrow{\mathrm{Heliosphere}}$ & Modulated Flux & $\xrightarrow{\mathrm{Magnetosphere}}$ & Primaries & $\cdots \xrightarrow{} $ \\
-& $\cdots \xrightarrow{}$ & $\xrightarrow{\mathrm{Atmosphere}}$ & Secondaries & $\xrightarrow{\mathrm{Detector}}$ & Signals \\
\end{tabular}
\end{table}
\end{center}

The above simulation scheme considers three essential factors with different spatial and time scales: the geomagnetic effects, the development of the extensive air showers in the atmosphere, and the detector response at ground level~\cite{AsoreyEtal2016a}. 

To take into account all these elements, we use ARTI: a complete computational framework for estimating the signals expected at the WCD detector and involving three main simulation blocks.~\cite{CalderonardilaEtal2019, SarmientocanoEtal2020}:
\begin{enumerate}
    \item The effects of the Geomagnetic Field (GF) on the propagation of charged particles, contributing to the background radiation at ground level, is characterized by the directional rigidity cutoff, $R_\mathrm{C}$.  This is performed at each LAGO site, using the MAGNETOCOSMICS code~\cite{Desorgher2004}, applying the backtracking technique~\cite{MasiasDasso2014}. The GF at any point on Earth is determined by using the International Geomagnetic Field Reference, version    11~\cite{IGRF11} at the near-Earth GF ($r<5R_{\oplus}$)\footnote{$r$ distance from Earth center and $R_{\oplus}$ is the Earth radius ($6371$\,km).} and through the Tsyganenko Magnetic Field model version 2001 (TSY01 hereafter)~\cite{Tsyganenko2002} to describe the outer GF ($r > 5 R_{\oplus}$).

    \item The second block simulates the extensive air showers produced during the interaction of cosmic rays with the atmosphere, obtaining a comprehensive set of secondaries at ground level. This block uses the CORSIKA (currently v7.6400) code~\cite{HeckEtal1998}, compiled with the following options: QGSJET-II-04~\cite{Ostapchenko2011}, GHEISHA-2002, EGS4, curved and external atmosphere, and volumetric detector.  
  
    \item Finally, with GEANT4, we model~\cite{Agostinelli2003} the detector response to the different types of impinging secondary particles obtaining the distribution of photo-electrons for a particular type of detector. The response of the WCD detectors is estimated by considering the detector geometry, e.g. cylindrical containers of water with an inner coating made of  Tyvek{\textregistered}~\cite{filevich1999spectral}, and a single photo-multiplier tube (PMT,  Hamamatsu R5912) located at the centre and top of the cylinder~\cite{Allard_etal2008}. 
\end{enumerate}
For simplicity, we have presented the above blocks as consecutive, but we sketched the precise ARTI operational pseudo-code in the Appendix. For a more detailed description, we refer the interested reader to the original paper~\cite{SarmientocanoEtal2020}.

The first two blocks have been integrated into a dedicated Virtual Organization, {\textit{lagoproject}}, as part of the European Grid Infrastructure (EGI, http://www.egi.eu) activities. The Grid implementation of CORSIKA was deployed with two ``flavours'', which run using GridWay Metascheduler (http://www.gridway.org/doku.php)~\cite{HuedoMonteroLlorente2001}. Massive calculations can be executed with the former, via the Montera~\cite{RodriguezMayoLlorente2013}, the GWpilot~\cite{RubioEtal2015}, the GWcloud~\cite{RubioHuedoMayo2015} and recently the EOSC-Synergy cloud services \cite{RubioEtal2021} frameworks.
 
\section{Atmospheric models with GDAS for the background study of secondary}
\label{sec:GDAS}
Understanding the propagation of EAS is decisive in estimating the flux of secondary particles at the detectors on the surface of the Earth. Therefore we must accurately characterize the atmosphere to simulate correctly the corresponding processes involved.  One of the essential atmosphere parameters is density, which determines the probability of interaction as the EAS evolves. The atmospheric density is concentrated in the first $30$~km from the ground up, decreasing as the altitude increases.

In this case, the atmospheric density is modelled by the vertical depth~\cite{Spurio2014}, measured in g/cm$^2$ and defined as

\begin{equation}
X_{h}= \int_{h}^{\infty} \rho (h') dh',
\label{eq:eq2}
\end{equation}
where $\rho(h)$ is the density as a function of height, $h$, above the Earth.  

From the ground up, five density layers model describe the atmosphere (i.e. Lindsley's standard atmospheric model~\cite{LinseyModel}), and an exponential can approximate the first four:

\begin{equation}
X_{h} = a_{i} + b_{i}e^{\frac{-h}{c_{i}}} \qquad \ i=1,..,4 \, .
\label{eq:eq25}
\end{equation}
At the same time the vertical atmospheric depth decreases linearly with height as

\begin{equation}
X_{h} = \alpha_{s}-\beta_{s}\frac{h}{\eta_{s}}\,;
\label{eq:eq26}
\end{equation}
where $a_{i}$, $b_{i}$, $c_{i}$, $\alpha_{s}$, $\beta_{s}$ and $\eta_{s}$ are the corresponding parameters of each atmospheric layer, which should be continuous across the boundaries of the different atmospheric segments~\cite{HeckEtal1998, PierreAugerObservatory2012}.

\subsection{EAS, CORSIKA and GDAS}
\label{EASCORSIKAGDAS}

CORSIKA applies Monte Carlo algorithms for EAS simulations and recreates its propagation when initiated by protons, photons, nuclei, or any other incoming particles~\cite{HeckEtal1998}. CORSIKA models the atmosphere through different types of configurations with a certain level of detail: 
\begin{itemize}
    \item ATMOD establishes predefined atmospheric models for specific locations, given the values of the parameters for each atmospheric layer.
    \item ATMEXT is a configuration for external atmospheres dependent on the geographical location, with the most common implemented models: tropical, mid-latitude summer, mid-latitude winter, sub-arctic summer, sub-arctic winter and U.S standard atmosphere at the South pole.
    \item Finally, the ATMFILE configuration lets us input a GDAS profile file previously created using the GDASTOOL code available in the CORSIKA software.
\end{itemize}

The configuration implemented for this work is ATMFILE Datacard using GDASTOOL~\cite{HeckEtal1998}, which allow us to create an atmospheric model from GDAS.  The GDAS atmospheric model incorporates the behaviour of the atmosphere based on meteorological observations (National Archive and Distribution System for Operational Models, NOAA). 

GDAS builds realistic climate predictions, describing the state of the atmosphere for certain variables in time and altitude. At a given time, $ t_{0}$, the observations give a value for a state variable, and at the same time, a forecast is available. The analysis stage combines observation and prediction to improve the model at $ t_{0}$. After curve fitting, a forecast is made for a later time $t_{1}$~\cite{PierreAugerObservatory2012}.

However, it is not enough to have an atmospheric profile for a given day and time to estimate the averaged secondary flux over any geographic location. A more robust model needs to be built, containing sufficient climate information over a given time interval. In this regard, previous studies have demonstrated the utility of building monthly GDAS atmospheric profiles on the CORSIKA's simulations~\cite{GrisalesSarmientoNunez2021}. 

Therefore, in this work, we propose a methodology that uses CORSIKA's GDASTOOL to extract an atmospheric profile for a specific day and time and build monthly average atmospheric profiles. GDASTOOL extracts values for the pressure, altitude, temperature and humidity, from GDAS-NOAA databases then fit them into the five-layer model implemented in CORSIKA~\cite{HeckEtal1998,PierreAugerObservatory2012}.

\subsection{Monthly atmospheric profiles}
 
\begin{figure}[!ht]
\centering
\includegraphics[width=0.7\textwidth]{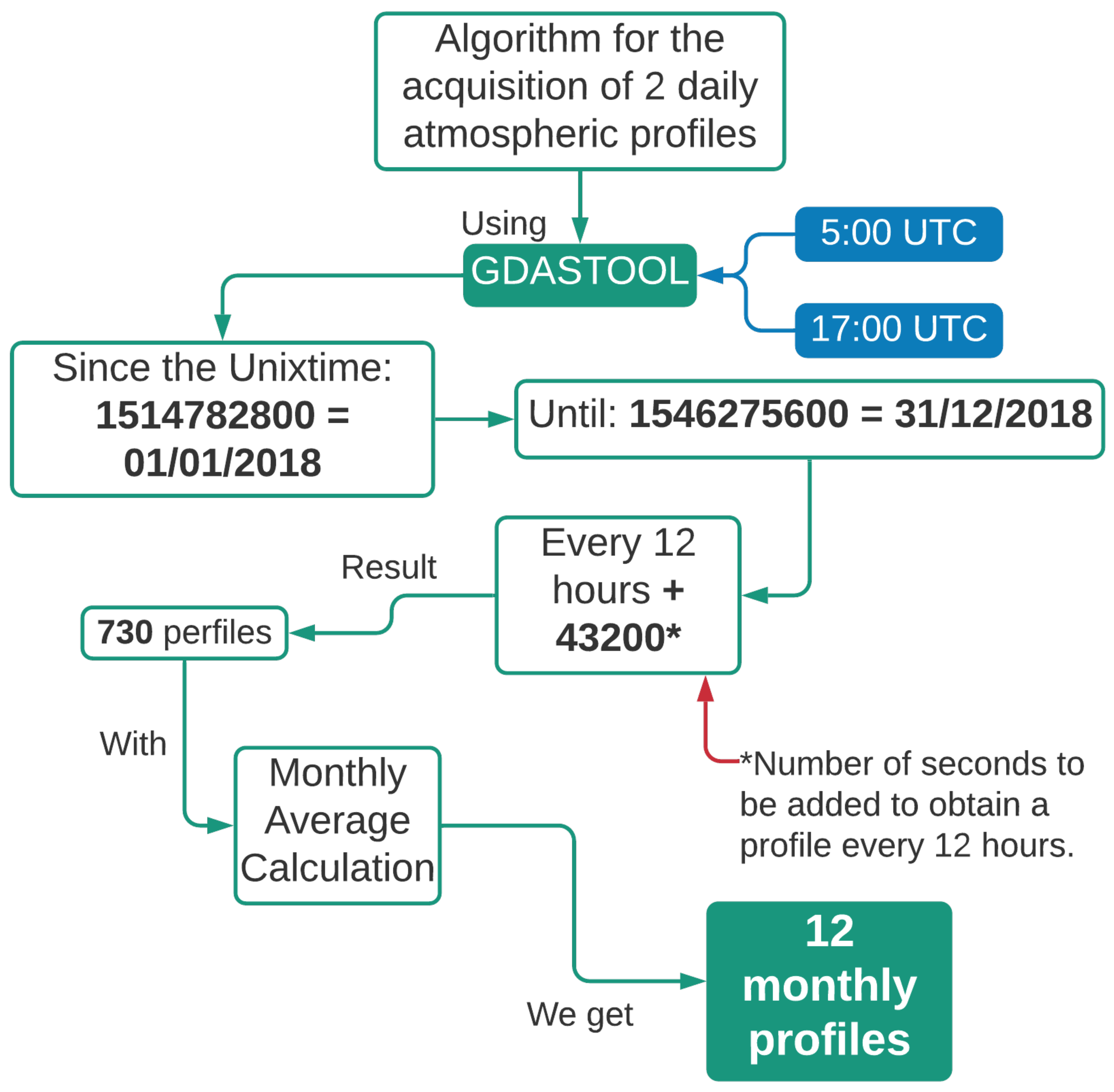}
\caption{The logical sequence is used to extract and build the 12-month average profiles for the city of Bucaramanga for 2018.}
\label{figLogicFlow}
\end{figure}

To build monthly atmospheric profiles, we implemented a computational algorithm using GDASTOOL that extracts data twice daily: at 0:00h and 12:00h, for all days of a year at any geographic position. In our case, as displayed in figure \ref{figLogicFlow}, it is for the city of Bucaramanga-Colombia (7.13$^{\circ}$~N, 73.00$^{\circ}$~W). 

\begin{figure}
\centering
\includegraphics[width=0.5\textwidth]{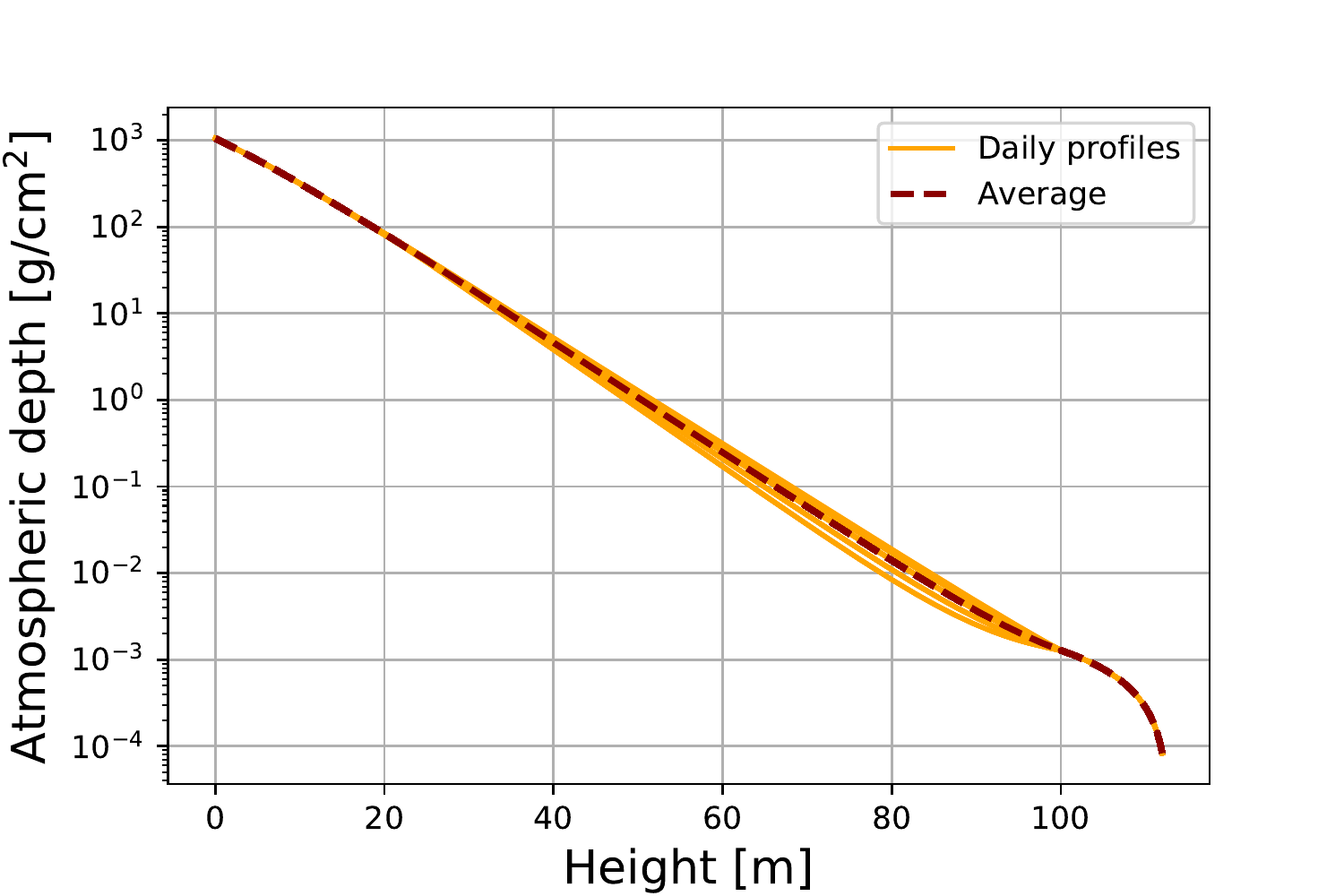}
\includegraphics[width=0.45\textwidth]{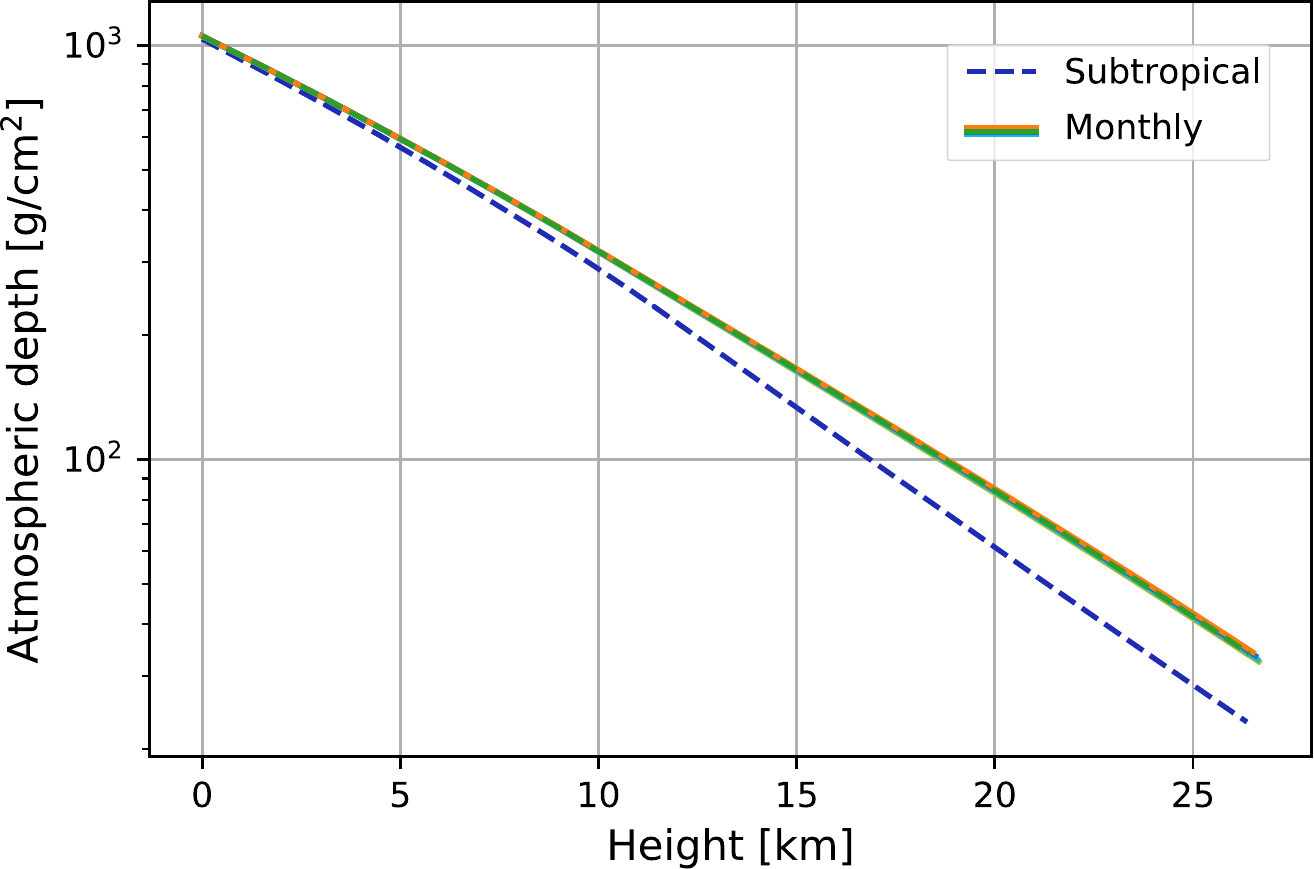}
\caption{Left panel shows as solid lines $62$ density GDAS profiles for January at Bucaramanga (7.13$^{\circ}$~N, 73.00$^{\circ}$~W) and the in-dash line, their average. The right side illustrates the first 30~km of the GDAS month density profiles for Bucaramanga and CORSIKA-tropical summer default profiles.}
\label{figMonthProfiles}
\end{figure}

We extracted 730 profiles per year, i.e. two profiles per day. Figure \ref{figMonthProfiles} (left plate) shows all the instantaneous profiles for January 2018 (solid line) and their average (dash line). As can be seen, some differences appear between them, although they highly agree in the initial $30$~km.

Thus, we built 12 monthly profiles for the year 2018 in the city Bucaramanga-Colombia and compared them with the predetermined profiles for this location (tropical summer/mid-latitude summer). 
Then, the new monthly atmospheric profile are the point-to-point average across the atmosphere over 60 daily profiles.
We observed significant differences, as shown on the right plate of figure \ref{figMonthProfiles}. Here, we plot the first 30~km, which accounts for most of the atmospheric matter density.

\subsection{Validating the method}

After obtaining the average atmospheric profiles, we must check if they reproduce the behaviour of the atmosphere in the middle of an EAS. Thus, we should apply this methodology to a location where the accuracy of GDAS was already known.

The selected location was the village of Malarg\"ue-Argentina, where the Pierre Auger Observatory is situated. This Observatory compared the GDAS data with local measurements (atmospheric soundings with weather balloons and ground-based weather stations), validating the accuracy of GDAS for the horizontal and vertical as well as temporal resolution\cite{PierreAugerObservatory2012}.

To validate our models, we built atmospheric profiles for Malarg\"ue between 2006 and 2011, extracting 10-month profiles every year on April the 6th, 12th, 18th, 24th and 30th, two per day, at 0:00h and 12:00h every day.

We then compare the evolution of the EAS using our GDAS model and the Auger atmospheric model, available as a predefined option in CORSIKA. We performed 100 EAS simulations for Iron primaries of $1~\times~10^{8}$~GeV. We made this choice due to two facts: the energy value should be in the maximum efficiency range, and the computation time of the simulation should not exceed one week of wall-clock time in our computational system.

\begin{figure}
\centering
\includegraphics[width=0.6\textwidth]{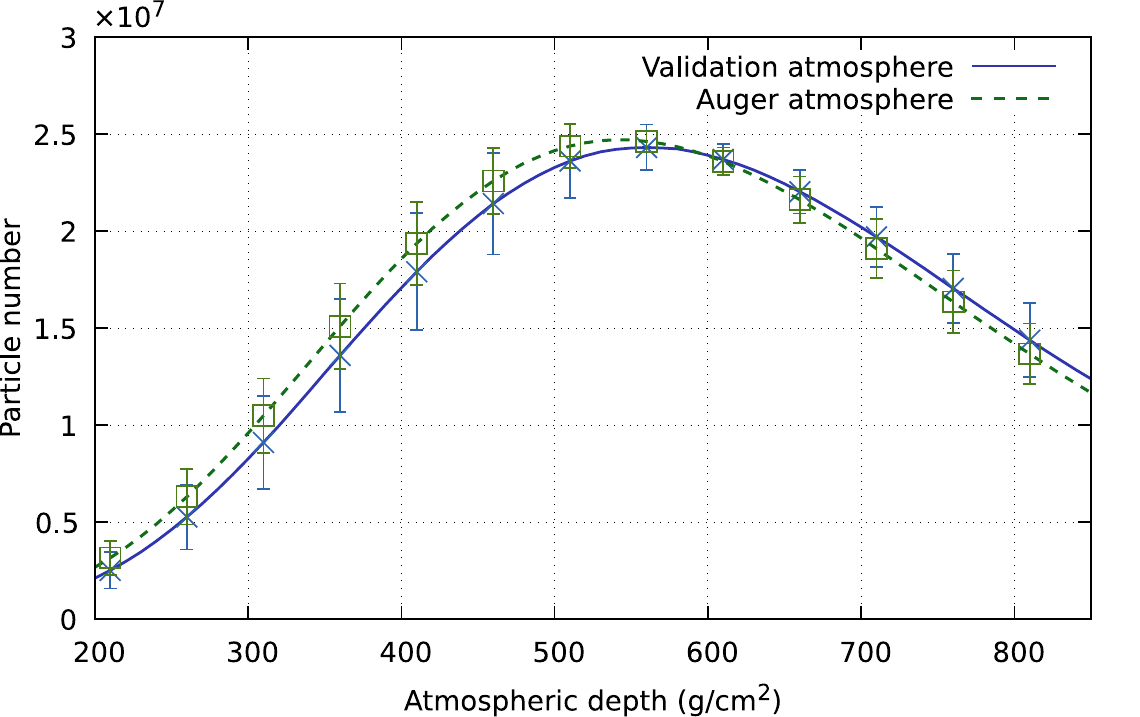}
\caption{Longitudinal distribution of secondary particles resulting from the interaction of an iron nucleus of $1~\times~10^{8}$~GeV over the atmosphere of Malarg\"ue for April. The Pierre Auger Observatory model in green and the atmosphere in the same conditions built by the methodology implemented for this work (Blue).  As it can be seen, the differences in this parameter did not exceed $2\%$ on both values of $X_{max}$.}
\label{figAugerGDAS}
\end{figure}

From the simulations and the analysis of the longitudinal development of the EAS, we identify the X$_\mathrm{max}$  corresponding to the maximum value of atmospheric depth, i.e. where the number of secondary particles is maximum. The X$_\mathrm{max}$  is a crucial parameter because it is proportional to the logarithm of the mass of the primary that started the EAS~\cite{Auger2014}. We have validated our methodology, by checking if the simulations yield a value of X$_\mathrm{max}$ close to those obtained by Pierre Auger Observatory profile. As it can be seen in figure \ref{figAugerGDAS}, the differences in this parameter did not exceed 2$\%$.

\section{Particle flux and atmospheric models}
\label{sec:FluxDependence}
First, we established the time needed to integrate the primary flux. For example, in the case of a 120-second flux, we calculated the total number of incident primaries and simulated each individual shower generated in the atmosphere.  Table \ref{PrimaryDist} (Appendix), shows the various primary particle contributing to the flux. The distribution corresponds to the abundances of the atomic nuclei reported in the literature~\cite{Spurio2014}.

Then we defined the initial conditions in order to run a series of simulations, which in our case, for the city of Bucaramanga, were established as:

\begin{itemize}
    \item Horizontal and vertical components of the Earth's magnetic field corresponding to 27.026~nT and 17.176~nT, respectively.
    \item Observation level, 950 m a.s.l. for Bucaramanga.
    \item Primary: Nuclei from Hydrogen to Iron
    \item Energy range of primaries: from 5~GeV to 10$^{6}$~GeV.
    \item Zenit angle of incidence of the primaries: from 0$^{\circ}$ to 90$^{\circ}$.
    \item Flux time 1~day = 86400~s.
    \item Type of detection: Volumetric.
    \item Atmospheric profile: Default tropical (mid-latitude summer) profile within ATMEXT routines, which is used so far for flux simulations over Bucaramanga, and the 12 monthly atmospheric profiles created from GDASTOOL.
    \item Energy cuts: E$_s$\,$\geq$\,5\,MeV for $\mu^\pm$s and hadrons, and E$_s$\,$\geq$\,5$\times$10$^{-2}$\,MeV for electrons and photons. Where E$_s$ is is the energy of secondary. These are the lower limits of the energies in this CORSIKA version.
\end{itemize}

\begin{figure}
\centering
\includegraphics[width=0.6\textwidth]{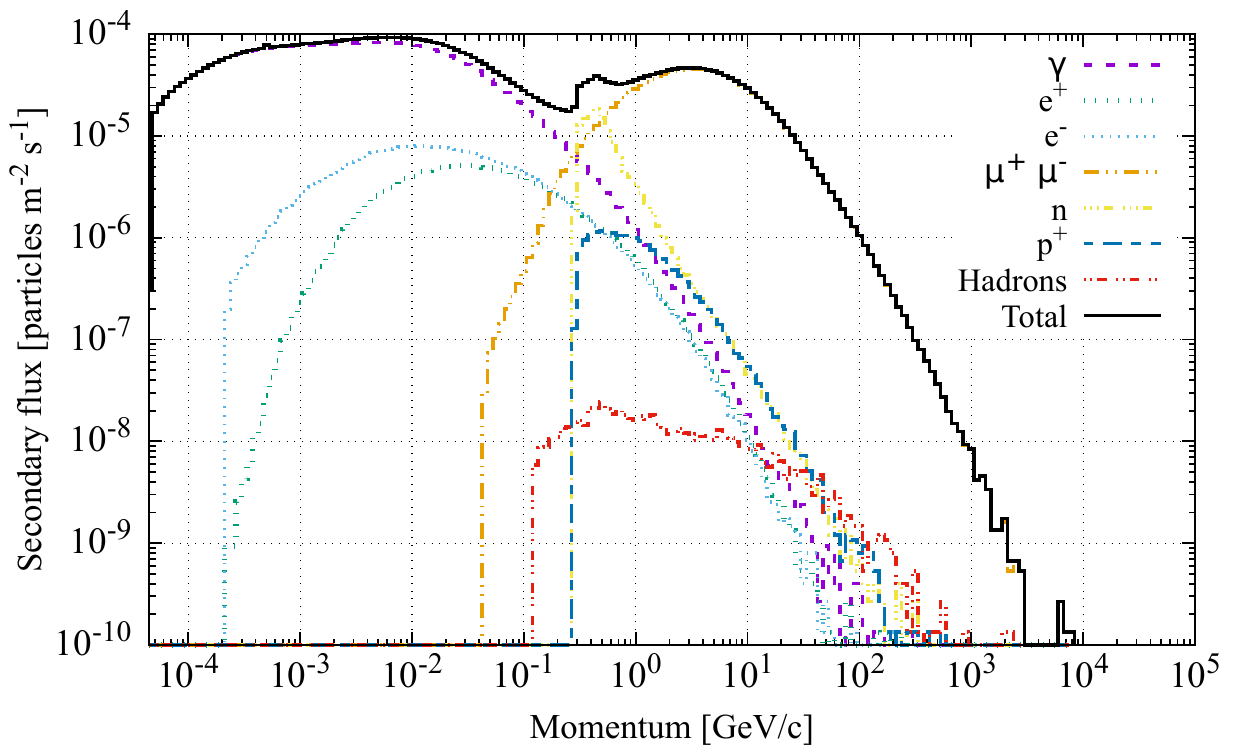}
\caption{Simulation of the energy spectrum of secondaries, at the level of Bucaramanga, using the atmospheric profile of April. The solid line represents the total secondary spectrum, and the dashed lines represent the contribution of photons, electrons, positrons, muons, neutrons, and protons separately.}
\label{fig:fig20}
\end{figure}

Figure \ref{fig:fig20} displays the spectrum of secondaries and the total secondary flux, using the constructed April atmospheric profile.  The neutron portion of the second hump is only significant between $0.2$~GeV/c and $1$~GeV/c, decreasing dramatically as the energy increases. In opposition, the muonic component increases in the same energy range, having its maximum value near $10$~GeV/c.

The plot shows two humps for each curve representing the secondary particle flux. The first hump represents the electromagnetic component (electrons, positrons and photons), while the second, made up of two smaller humps, represents the flux of neutrons and muons, respectively.

We ran a total of $12$ flux simulations using a GDAS monthly atmospheric profile at a time and, finally, one simulation using the CORSIKA tropical summer (mid-latitude summer) predefined profile available in ATMEXT configuration.

\begin{figure}
\centering
\includegraphics[width=1\textwidth]{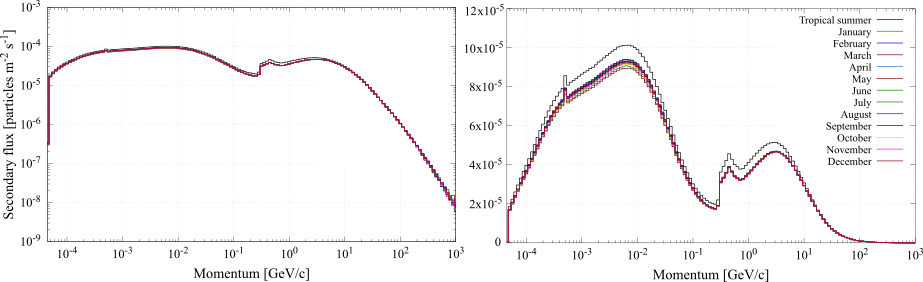}
\caption{Simulation of total secondary flux as a function of energy at Bucaramanga, using different means of interaction. The flux is in a logarithmic scale (left) and linear (right). In both cases, the black line represents the tropical summer profile, and the others correspond to the $12$ monthly atmospheric profiles. The calculations show a overestimation of the tropical profile compared to the 12 GDAS atmospheres.}
\label{fig:fig199}
\end{figure}

Figure \ref{fig:fig199} shows the total secondary flux as a function of energy at the altitude of Bucaramanga, using different atmospheric models. The black line corresponds to the simulation using the default tropical predefined profile, and the coloured lines correspond to the $12$ monthly GDAS atmospheric models. As can be appreciated, there is a higher flux with the predefined atmosphere compared to the $12$ monthly profiles.

\section{Final remarks}
\label{sec:FinalRemarks}

We have devised a methodology that enables us to obtain a month-by-month averaged atmospheric profile for any geographic location. This methodology, implemented using the GDASTOOL code,  extracts meteorological data for Bucaramanga at noon and midnight: (0:00h and 12:00 (UTC-5)), during a whole year. In this way, we created $12$ atmospheric profiles for 2018 and compared them with predefined atmospheric profiles available in CORSIKA.

We validated this methodology, by generating atmospheric profiles for the Pierre Auger Observatory and contrasting them with the Observatory's GDAS-based models. The behaviour of the EAS obtained with the reconstructed atmosphere shows a difference of $\approx 2\%$ in the value of the maximum atmospheric depth, X$_\mathrm{max}$.

It is essential to clarify that monthly atmospheric profiles over a year are not sufficient to represent the average climatic variability in the tropics. We need atmospheres that cover a more significant number of years. Thus, the observed differences suggest the importance of studying these effects in greater detail, for example, long-term climatic events such as the El Ni\~no and La Ni\~na phenomena.

This work completes the sequence of simulations that enables to study, in a realistic approach, the flux of secondaries that a WCD can detect at any geographical position and time of the year. 

\section*{Acknowledgements}

The authors are grateful to Dr. Hern\'an Asorey for his constructive comments on previous versions of the manuscript. We also thank the Pierre Auger Observatory Collaboration members for their continuous engagement and support. LAN gratefully acknowledge the permanent support of Vicerrector\'ia de Investigaci\'on y Extensi\'on de la Universidad Industrial de Santander. This work has been partially carried out on the ACME cluster, which is owned by CIEMAT and funded by the Spanish Ministry of Science and Innovation project CODEC-OSE (RTI2018-096006-B-I00) with FEDER funds as well as supported by the CYTED co-founded RICAP Network (517RT0529). We also acknowledge the computational support from the Universidad Industrial de Santander (SC3UIS) High Performance and Scientific Computing Centre.


\newpage

\begin{table}
\small\sf\centering
\caption{Distribution of primaries to be simulated for a secondary flow of $120$~s at the height of Bucaramanga obtained by ARTI.}
\begin{tabular}{llll}
\toprule
\multicolumn{1}{c}{\textbf{Nuclei}} & \multicolumn{1}{c}{\textbf{Quantity}} & \textbf{Nuclei} & \textbf{Quantity} \\
\midrule
H & 562322 & Al & 44 \\ \hline
He & 56595 & Na & 38 \\ \hline
C & 1458 & Ca & 30 \\ \hline
O & 1410 & F & 25 \\ \hline
Li & 574 & Cr & 19 \\ \hline
B & 396 & Ar & 18 \\ \hline
Mg & 335 & Ti & 17 \\ \hline
Si & 322 & Mn & 13 \\ \hline
N & 295 & K & 11 \\ \hline
Ne & 259 & V & 10 \\ \hline
Fe & 195 & P & 9 \\ \hline
Be & 167 & Cl & 8 \\ \hline
S & 51 & Sc & 5 \\ \hline
\bottomrule
\end{tabular}
\label{PrimaryDist}
\end{table}

\appendix
\section{Appendix}

The following algorithm represents the three main parts that make up ARTI for flux simulations at CORSIKA, magnetic field correction via Magnetocosmic and detector simulation via Geant4 (See reference~\cite{SarmientocanoEtal2020} for a more detailed description).
\begin{algorithm}
    \SetKwInOut{Input}{Input}
    \SetKwInOut{Output}{Output}

    \underline{Simulation flux}\;
    \Input{E = [1, 10$^6$] GeV; energy range\\
    injected primary nuclei, Z = [1, 26] \\
	$\theta$ = [0, 90]; zenith angle range\\
	$\phi$ = [-180, 180]; azimuth angle range\\
	Altitude in m a.s.l. \\
	B$_x$, B$_z$; site's magnetic field\\
	GDAS model; atmospheric model\\
	time; 86400 sec in this case\\}
    \Output{$\Xi$, particle flux at ground}
    \Begin{
    $\Phi(E_p, Z, A, \Omega)$; Integrate astroparticles spectra\\
	Z, \#part(E) $\rightarrow$ built steering Corsika files\\
	run block, via Corsika software\\
	Analisys block; read and uncompress binary files\\
	}
      {
        return $\Xi$\;
      }
      
    \underline{Magnetic field correction}\;   
    \Input{$\Xi$, particle flux at ground\\
	Rm, magnetic rigidity\\
	IGRF, magnetic model\\}
    \Output{$\Xi_{corr}$, particle flux at ground corrected}  
    \Begin{
    $R_\mathrm{C}(\phi, \theta)$, magnetic rigidity cutoff\\
    }
      {
        return $\Xi_{corr}$\;
      }
    \underline{Magnetic field correction}\; 
    \Input{$\Xi_{corr}$, particle flux at ground corrected\\
	D(r,h), detector's dimensions\\
	PMT, photo-multiplier's features\\
	}
	\Output{E$_{D}$, energy deposited}
    \For{particles}
    {intercating with water\\
    E$^{D}_{i}$, energy deposited by i-particle}
      {
        return E$_{D}$\;
      } 
 
    \caption{ARTI is divided in three parts. Flux simulations via Corsika,
magnetic field corretion via Magnetocosmic and detector simulation via Geant4}
\end{algorithm}

\end{document}